\documentclass[12pt]{article}

\usepackage{algorithm}
\usepackage{algpseudocode}
\usepackage{amsmath}
\usepackage{amssymb}
\usepackage{amsthm}
\usepackage{bm}
\usepackage{booktabs}
\usepackage[bf,footnotesize]{caption}
\usepackage{fullpage}
\usepackage{graphicx}
\usepackage[sc]{mathpazo}
\usepackage{mathtools}
\usepackage[round]{natbib}
\usepackage{xcolor}

\algrenewcommand\algorithmicrequire{\textbf{Input:}}
\algrenewcommand\algorithmicensure{\textbf{Output:}}

\newtheorem{theorem}{Theorem}

\newcommand{\T}{\intercal}
\newcommand{\eq}[1]{Eq.~(\ref{eq:#1})}
\newcommand{\sect}[1]{Section~\ref{sec:#1}}

\allowdisplaybreaks

\title{\bf Meeting times on graphs in near-cubic time}
\author{Alex McAvoy \\ \small School of Data and Information Sciences \\
	\small Department of Mathematics \\
	\small University of North Carolina at Chapel Hill}
\date{}

\begin{document}

\maketitle

\begin{abstract}
The expected meeting time of two random walkers on an undirected graph of size $N$, where at each time step one walker moves and the process stops when they collide (the ``lazy'' pair walk), satisfies a system of $\binom{N}{2}$ linear equations. Na\"{i}vely, solving this system takes $O\left(N^{6}\right)$ operations. However, this system of linear equations has nice structure in that it is \emph{almost} a Sylvester equation, with the obstruction being a diagonal absorption constraint. We give a simple algorithm for solving this system that exploits this structure, leading to $O\left(N^{4}\right)$ operations and $\Theta\left(N^{2}\right)$ space for direct computation of all $\binom{N}{2}$ meeting times. While this practical method uses only standard dense linear algebra, it can be improved to $O\left(N^{3}\log^{2}N\right)$ operations by exploiting the Cauchy structure of the diagonal correction. We treat the near-cubic bound as theoretical, however, since it appears to involve a large constant factor. In practice, the dense $O\left(N^{4}\right)$ algorithm computes all meeting times on graphs with $N$ in the low thousands in minutes on a laptop. We generalize this result slightly to cover the Poisson equation for the absorbing lazy pair walk with an arbitrary source, which can be solved at the same cost, with $O\left(N^{3}\right)$ per additional source on the same graph. We conclude with applications to evolutionary dynamics, giving improved algorithms for calculating fixation probabilities and mean trait frequencies.
\end{abstract}

\section{Introduction}
Expected meeting times of two random walkers on a graph arise in Markov chain theory \citep{aldous2002reversible} and evolutionary dynamics on graphs \citep{allen:Nature:2017,mcavoy:JMB:2021}. Let $W$ be the adjacency matrix of a connected, undirected, weighted graph on $N$ nodes, and let $P_{ij}\coloneqq W_{ij}/w_{i}$ (with $w_{i}\coloneqq\sum_{k=1}^{N}W_{ik}$) be the transition probabilities of the associated random walk. The lazy pair walk places two walkers on the graph; at each step, one walker (chosen uniformly) moves, and the process absorbs when they collide. It is ``lazy'' in the sense that one walker rests while the other moves. The laziness assumption is essential for what follows because if both walkers move simultaneously, then the Sylvester structure exploited below is lost (see \sect{discussion}).

The expected meeting time $\tau_{ij}$ from starting positions $\left(i,j\right)$ satisfies the recurrence
\begin{align}
\tau_{ij} &= \begin{cases}
0 & i=j , \\
1 + \frac{1}{2}\sum_{k=1}^{N}P_{ik}\tau_{kj}+\frac{1}{2}\sum_{k=1}^{N}P_{jk}\tau_{ik} & i\neq j ,
\end{cases}\label{eq:meeting}
\end{align}
which is a linear system of $\binom{N}{2}$ equations in $\binom{N}{2}$ unknowns. \eq{meeting} is a special case of the Poisson equation for the absorbing (lazy) pair walk with symmetric source, $F$, and prescribed diagonal, $h=\left(h_{1},\dots ,h_{N}\right)$, defined by
\begin{align}
H_{ij} &= \begin{cases}
h_{i} & i=j , \\
F_{ij} + \frac{1}{2}\sum_{k=1}^{N}P_{ik}H_{kj} + \frac{1}{2}\sum_{k=1}^{N}P_{jk}H_{ik} & i\neq j .
\end{cases}\label{eq:poisson}
\end{align}
\eq{poisson} describes a Sylvester equation with a diagonal constraint. The key observation is that a standard Sylvester equation $X - \frac{1}{2}PX - \frac{1}{2}XP^{\T} = B$ can be solved in $O\left(N^{3}\right)$ by the Bartels-Stewart algorithm \citep{bartels:CACM:1972,golub:TAC:1979}, because the eigenbasis of $P$ diagonalizes $P\otimes I+I\otimes P^{\T}$. The diagonal constraint is the sole obstruction to applying this directly.

For a single walker, commute and hitting times are computable in $O\left(N^{3}\right)$ via the Laplacian pseudoinverse \citep{tetali:JAP:1991,chandra:CC:1996,beveridge:CPC:2015}. For two walkers, the pair system has $\binom{N}{2}$ terms, and na\"{i}ve Gaussian elimination applied to this system requires $O\left(N^{6}\right)$ operations; the best previous exact bound is $O\left(N^{4.75}\right)$ \citep{coppersmith:JSC:1990}, improved slightly to $O\left(N^{4.74}\right)$ by \citet{williams:SODA:2024}. Of course, the latter refers to a bound on generic linear systems with $N^{2}$ unknowns and does not, itself, exploit the structure of meeting times. In practice, most existing approaches use either sparse solvers or iterative methods that depend on graph density.

We show that the diagonal constraint can be resolved efficiently using the Cauchy structure of the eigenbasis divisors, yielding a parameter-free algorithm at $O\left(N^{3}\log^{2}N\right)$ operations in theory and $O\left(N^{4}\right)$ in practice. The same algorithm solves \eq{poisson} for any symmetric source $F$, with $O\left(N^{3}\right)$ per additional source after a one-time ``source-free'' setup. Perhaps just as importantly, this algorithm requires $\Theta\left(N^{2}\right)$ space, which is asymptotically optimal in the Poisson equation for the absorbing pair walk. In practice, the $O\left(N^{4}\right)$ variant computes all meeting times on the empirical networks of \sect{evol} (for $N$ up to $2{,}312$) in seconds to minutes on a laptop, regardless of network density (\sect{benchmarks}). This graph-dependent computation is needed only once; under weak selection, every game on the same graph then requires $O\left(N^{3}\right)$ operations. We develop the algorithm in \sect{algorithm} and give the evolutionary dynamics applications in \sect{evol}.

\section{Algorithm}\label{sec:algorithm}

\subsection{Diagonal extension and symmetrization}\label{sec:diag_extend}
To obtain a simple matrix equation from \eq{poisson} that holds on the diagonal as well, we can introduce a diagonal slack $D = \mathrm{diag}\left(d_{1},\dots ,d_{N}\right)$ and write, for all entries,
\begin{align}
H - \frac{1}{2}PH - \frac{1}{2}HP^{\T} &= F + D , \label{eq:full_sylvester}
\end{align}
subject to $H_{ii}=h_{i}$ for $i=1,\dots ,N$. Eq.~\eqref{eq:full_sylvester} is now a standard Sylvester equation, with the matrix $D$ being an unknown quantity.

The random walk is reversible, which means $\pi_{i}P_{ij}=\pi_{j}P_{ji}$ for all $i,j=1,\dots ,N$, where $\pi$ is the stationary distribution for the walk (in this case, $\pi_{i}$ is proportional to $w_{i}$). Let $\Pi\coloneqq\mathrm{diag}\left(\pi_{1},\dots ,\pi_{N}\right)$ and consider the matrices $\widetilde{P}\coloneqq\Pi^{1/2}P\Pi^{-1/2}$, $\widehat{H}\coloneqq\Pi^{1/2}H\Pi^{1/2}$, $\widehat{F}\coloneqq\Pi^{1/2}F\Pi^{1/2}$, and $\widehat{D}\coloneqq\Pi^{1/2}D\Pi^{1/2}=\Pi D$. From \eq{full_sylvester}, we then obtain
\begin{align}
\widehat{H} - \frac{1}{2}\widetilde{P}\widehat{H} - \frac{1}{2}\widehat{H}\widetilde{P} &= \widehat{F} + \widehat{D} , \label{eq:sylvester_sym}
\end{align}
which is a Sylvester equation involving only symmetric matrices.

Since $\widehat{D}_{ij}=0$ when $i\neq j$ and $\widehat{H}_{ii}=\pi_{i}h_{i}$, $\widehat{F}_{ii}=\pi_{i}F_{ii}$, and $\widehat{D}_{ii}=\pi_{i}d_{i}$, we have
\begin{align}
\pi_{i}d_{i} &= \pi_{i}h_{i} - \frac{1}{2}\left(\widetilde{P}\widehat{H}\right)_{ii} - \frac{1}{2}\left(\widehat{H}\widetilde{P}\right)_{ii} - \pi_{i}F_{ii} . \label{eq:diagonal_constraint}
\end{align}
We use \eq{diagonal_constraint} later in deriving a linear system that uniquely determines $d$.

\subsection{Eigenbasis diagonalization}\label{sec:eigenbasis}
Write $\widetilde{P} = V\Lambda V^{\T}$ with $V$ orthonormal and $\Lambda = \mathrm{diag}\left(\lambda_{1},\dots ,\lambda_{N}\right)$. The eigenvalues of $\widetilde{P}$ are the same as those of $P$ and can be ordered such that $\lambda_{1}=1\geqslant\lambda_{2}\geqslant\cdots\geqslant\lambda_{N}\geqslant -1$ since $P$ is stochastic. Moreover, by the Perron-Frobenius theorem, the eigenspace associated to $\lambda_{1}=1$ has dimension one. Conjugating \eq{sylvester_sym} by $V^{\T}$ gives the equation
\begin{align}
\widehat{H}^{V} - \frac{1}{2}\Lambda\widehat{H}^{V} - \frac{1}{2}\widehat{H}^{V}\Lambda &= \widehat{F}^{V} + \widehat{D}^{V} ,
\end{align}
where $\widehat{H}^{V}\coloneqq V^{\T}\widehat{H}V$, $\widehat{F}^{V}\coloneqq V^{\T}\widehat{F}V$, and $\widehat{D}^{V}\coloneqq V^{\T}\widehat{D}V$. Therefore, for every $i,j=1,\dots ,N$,
\begin{align*}
\left(1-\frac{1}{2}\lambda_{i}-\frac{1}{2}\lambda_{j}\right)\widehat{H}_{ij}^{V} &= \widehat{F}_{ij}^{V} + \widehat{D}_{ij}^{V} .
\end{align*}
Since $\lambda_{1}=1$ is a simple eigenvalue for $\widetilde{P}$, we have $\widehat{F}_{11}^{V}+\widehat{D}_{11}^{V}=0$ and
\begin{align}
\widehat{H}_{ij}^{V} &= \frac{\widehat{F}_{ij}^{V}+\widehat{D}_{ij}^{V}}{1-\frac{1}{2}\lambda_{i}-\frac{1}{2}\lambda_{j}} \label{eq:HV_entry}
\end{align}
for all $\left(i,j\right)\neq\left(1,1\right)$. Here, we note that $\widehat{H}_{11}^{V}$ is ``free'' for the moment and still unknown.

Without the diagonal constraint, \eq{HV_entry} would be the standard Bartels-Stewart algorithm \citep{bartels:CACM:1972}, which in this case requires $O\left(N^{3}\right)$ operations. The constraint is the source of the additional complexity, which we resolve next using \eq{diagonal_constraint}.

\subsection{Diagonal correction}\label{sec:diag_resolve}
Evaluating \eq{sylvester_sym} along the diagonal gives $\left(\widetilde{P}\widehat{H}\right)_{ii}+\left(\widehat{H}\widetilde{P}\right)_{ii}=2\pi_{i}h_{i}-2\pi_{i}F_{ii}-2\pi_{i}d_{i}$. Expanding the left-hand side using $\widetilde{P}\widehat{H}=V\Lambda\widehat{H}^{V}V^{\T}$ and $\widehat{H}\widetilde{P}=V\widehat{H}^{V}\Lambda V^{\T}$, we obtain
\begin{align}
2\pi_{i}h_{i} - 2\pi_{i}F_{ii} - 2\pi_{i}d_{i} &= \sum_{j,k=1}^{N}V_{ij}\left(\lambda_{j}+\lambda_{k}\right)\widehat{H}_{jk}^{V}V_{ik} \nonumber \\
&= 2\pi_{i}\widehat{H}_{11}^{V} + \sum_{\left(j,k\right)\neq\left(1,1\right)}V_{ij}\frac{\lambda_{j}+\lambda_{k}}{1-\frac{1}{2}\lambda_{j}-\frac{1}{2}\lambda_{k}}\left(\widehat{F}_{jk}^{V}+\widehat{D}_{jk}^{V}\right)V_{ik} \nonumber \\
&= 2\pi_{i}\widehat{H}_{11}^{V} + 2\sum_{\left(j,k\right)\neq\left(1,1\right)}\frac{V_{ij}\left(\widehat{F}_{jk}^{V}+\widehat{D}_{jk}^{V}\right) V_{ik}}{1-\frac{1}{2}\lambda_{j}-\frac{1}{2}\lambda_{k}} \nonumber \\
&\quad - 2\sum_{\left(j,k\right)\neq\left(1,1\right)}V_{ij}\left(\widehat{F}_{jk}^{V}+\widehat{D}_{jk}^{V}\right) V_{ik}  \nonumber \\
&= 2\pi_{i}\widehat{H}_{11}^{V} + 2\sum_{\left(j,k\right)\neq\left(1,1\right)}\frac{V_{ij}\left(\widehat{F}_{jk}^{V}+\widehat{D}_{jk}^{V}\right) V_{ik}}{1-\frac{1}{2}\lambda_{j}-\frac{1}{2}\lambda_{k}} -2\pi_{i}F_{ii}-2\pi_{i}d_{i} \label{eq:diag_expand}
\end{align}
for all $i=1,\dots ,N$, since $\lambda_{1}=1$ and $V_{i1}=\sqrt{\pi_{i}}$. Therefore,
\begin{align}
\pi_{i}h_{i} &= \pi_{i}\widehat{H}_{11}^{V} + \sum_{\left(j,k\right)\neq\left(1,1\right)}\frac{V_{ij}\left(\widehat{F}_{jk}^{V}+\widehat{D}_{jk}^{V}\right) V_{ik}}{1-\frac{1}{2}\lambda_{j}-\frac{1}{2}\lambda_{k}} . \label{eq:correction_constraint_withoutM}
\end{align}

To write this in a slightly more palatable way, consider the matrix, $S$, with entries
\begin{align}
S_{jk} &\coloneqq
\begin{cases}
1 & \left(j,k\right) =\left(1,1\right) , \\
\frac{1}{\frac{1}{2}\left(1-\lambda_{j}\right) -\frac{1}{2}\left(\lambda_{k}-1\right)} & \left(j,k\right)\neq\left(1,1\right) .
\end{cases}
\end{align}
If we define the vector $f$ and matrix $M$ with entries
\begin{subequations}\label{eq:f_and_M}
\begin{align}
f_{i} &\coloneqq \sum_{j,k=1}^{N}V_{ij}\widehat{F}_{jk}^{V}S_{jk}V_{ik} ; \\
M_{ij} &\coloneqq \sum_{k,\ell =1}^{N} V_{ik}V_{jk}S_{k\ell}V_{j\ell}V_{i\ell} ,
\end{align}
\end{subequations}
then, with the condition $\widehat{F}_{11}^{V}+\widehat{D}_{11}^{V}=0$, \eq{correction_constraint_withoutM} yields the linear system
\begin{align}
\begin{pmatrix}
0 & \pi^{\T} \\
\pi & M
\end{pmatrix}
\begin{pmatrix}
\widehat{H}_{11}^{V} \\ \pi\odot d
\end{pmatrix}
&= 
\begin{pmatrix}
-\widehat{F}_{11}^{V} \\ \pi\odot h-f
\end{pmatrix} . \label{eq:correction_constraint}
\end{align}
With the matrix $U\coloneqq\left(V_{1\ast}\otimes V_{1\ast} \mid V_{2\ast}\otimes V_{2\ast} \mid \cdots \mid V_{N\ast}\otimes V_{N\ast}\right)$, we can write
\begin{align}
M &= U^{\T}\textrm{diag}\left(\textrm{vec}\left(S\right)\right) U ,
\end{align}
which has full rank because $S$ is positive and the columns of $V$ form an eigenbasis for $\widetilde{P}$. Therefore, \eq{correction_constraint} has a unique solution since $M$ is positive-definite and $\pi\neq 0$. Solving this equation yields values of $d_{1},\dots ,d_{N}$, from which we can form $\widehat{D}^{V}$ for use in \eq{HV_entry}. (Note that $M$ depends on the graph but not on the source or the diagonal constraint.)

\subsection{Cauchy structure and complexity}\label{sec:cauchy}
Forming the matrix $M$ of \eq{f_and_M} entry-by-entry requires $O\left(N^{4}\right)$ using standard (dense) linear algebra routines. However, since $M_{ij}=\sum_{k,\ell =1}^{N} \left(V_{i\ast}\otimes V_{j\ast}\right)_{kk} S_{k\ell} \left(V_{i\ast}\otimes V_{j\ast}\right)_{\ell\ell}$ for all $i,j=1,\dots ,N$ and $S$ is a Cauchy matrix, computing $M_{ij}$ requires the product of a Cauchy matrix and a vector, followed by an inner product. Products of Cauchy matrices and vectors use $O\left(N\log^{2}N\right)$ operations via FFT-based polynomial arithmetic \citep[see][Problem $3.6.1$]{pan:Birkhauser:2001}. Since this bound determines the theoretical complexity of computing each entry of $M$, constructing the entire matrix requires $O\left(N^{3}\log^{2}N\right)$ operations.

\subsection{The complete algorithm}\label{sec:alg_steps}
Putting all of these steps together, we obtain Algorithm~\ref{alg:meeting} and Theorem~\ref{thm:meeting}.

\begin{algorithm}[ht]
\caption{Poisson equation for the absorbing lazy pair walk}\label{alg:meeting}
\begin{algorithmic}[1]
\Require Transition matrix $P$, symmetric source $F$, and prescribed diagonal, $h$
\Ensure Solution $H$ of \eq{poisson}
\Statex \textit{One-time setup (depends on only the graph and not on the source):}
\State Eigendecompose $\widetilde{P} = \Pi^{1/2}P\Pi^{-1/2} = V\Lambda V^\T$ \hfill $O\left(N^{3}\right)$
\State Form matrix of \eq{f_and_M} using Cauchy matrix-vector products \hfill $O\left(N^{3}\log^{2}N\right)$
\Statex
\Statex \textit{Per-source ($F$) and prescribed diagonal ($h$) computation:}
\State Compute $\widehat{F}^V = V^\T\Pi^{1/2}F\Pi^{1/2}V$ \hfill $O\left(N^{3}\right)$
\State Solve \eq{correction_constraint} for $\widehat{H}_{11}^{V}$ and $d$, and then form $\widehat{D}^{V}$ \hfill $O\left(N^{3}\right)$
\State Set $\widehat{H}^V_{jk} =\left(\widehat{F}_{jk}^{V}+\widehat{D}_{jk}^{V}\right) /\left(1-\frac{1}{2}\lambda_{j}-\frac{1}{2}\lambda_{k}\right)$ for $\left(j,k\right)\neq\left(1,1\right)$ \hfill $O\left(N^{2}\right)$
\State Compute $H = \Pi^{-1/2}V\widehat{H}^{V}V^{\T}\Pi^{-1/2}$ \hfill $O\left(N^{3}\right)$
\end{algorithmic}
\end{algorithm}

\begin{theorem}\label{thm:meeting}
The Poisson equation for the lazy pair walk on any weighted, undirected, connected graph can be solved in $O\left(N^{3}\log^{2}N\right)$ operations and $\Theta\left(N^{2}\right)$ space. The time per additional source and diagonal constraint is $O\left(N^{3}\right)$, after the one-time setup of $O\left(N^{3}\log^{2}N\right)$ operations.
\end{theorem}

The theoretical bound of $O\left(N^{3}\log^{2}N\right)$ relies on fast Cauchy matrix-vector products, but the nature of that algorithm appears to involve a large constant factor. Forming $M$ entry-by-entry directly requires $O\left(N^{4}\right)$ operations but can take advantage of standard, optimized linear algebra routines. The time for the latter would surpass the former for large $N$, but in practice we recommend using the entry-by-entry method.

Algorithm~\ref{alg:meeting} works with dense $N\times N$ matrices and uses $\Theta\left(N^{2}\right)$ memory, which is asymptotically optimal for any algorithm whose output is the full matrix $H$, since $H$ itself has $\Theta\left(N^{2}\right)$ entries. Both the runtime ($O\left(N^{4}\right)$ practical and $O\left(N^{3}\log^{2}N\right)$ theoretical) and the memory of Algorithm~\ref{alg:meeting} are determined entirely by $N$. By contrast, solving \eq{poisson} via the full $N^{2}\times N^{2}$ system requires $\Theta\left(N^{4}\right)$ memory stored densely, or $\Theta\left(N\left|E\right|\right)$ in a CSR sparse format, both of which become infeasible on moderately dense graphs of moderate size (e.g., with $N$ in the hundreds).

\subsection{Benchmarks}\label{sec:benchmarks}
Table~\ref{tab:benchmarks} gives the running time and memory for Algorithm~\ref{alg:meeting} on Erd\H{o}s-R\'{e}nyi graphs and on several empirical networks (used later in \sect{evol}). All runs were carried out in double precision on a laptop (Apple M4, 10 cores, 16 GB RAM) with Python 3.13.9, NumPy 2.3.4, and SciPy 1.16.3, with Apple's Accelerate framework as the BLAS/LAPACK backend. Times are wall-clock times of individual runs, and memory reports the peak memory used in a fresh process for each graph.

\begin{table}[ht]
\centering
\caption{Wall-clock running times (seconds) and peak memory of the $O\left(N^{4}\right)$ implementation of Algorithm~\ref{alg:meeting} for computing all $\binom{N}{2}$ meeting times (i.e., $F=\mathbf{1}\mathbf{1}^{\T}$ and $h=0$). The stages are the eigendecomposition of $\widetilde{P}$ (``eig''); forming $M$ of \eq{f_and_M} entry-by-entry (``form $M$''); forming $\widehat{F}^{V}$ and $f$ and solving \eq{correction_constraint} (``solve''); and forming $\widehat{D}^{V}$ and $\widehat{H}^{V}$ and reconstructing $H$ (``reconstruct''). The first two stages are the one-time, source-independent setup, whereas the last two are repeated for each source. ``Total'' is the sum of the four stages. ``Memory'' is the peak memory of the Python process, including the interpreter and libraries. The Erd\H{o}s-R\'{e}nyi graphs have mean degree $40$ (largest connected component), the Facebook networks are those of Figure~\ref{fig:facebook_networks}, and the email network is that of Figure~\ref{fig:email_network}.}\label{tab:benchmarks}
\small\setlength{\tabcolsep}{4.5pt}
\begin{tabular}{lrrrrrrrr}
\toprule
Graph & $N$ & $\left|E\right|$ & eig & form $M$ & solve & reconstruct & total & memory (MB) \\
\midrule
Erd\H{o}s-R\'{e}nyi & $250$ & $4{,}962$ & $0.00$ & $0.0$ & $0.01$ & $0.01$ & $0.1$ & $78$ \\
Erd\H{o}s-R\'{e}nyi & $500$ & $10{,}107$ & $0.01$ & $0.4$ & $0.05$ & $0.09$ & $0.6$ & $111$ \\
Erd\H{o}s-R\'{e}nyi & $1{,}000$ & $19{,}853$ & $0.06$ & $5.7$ & $0.39$ & $0.82$ & $7.0$ & $256$ \\
Erd\H{o}s-R\'{e}nyi & $2{,}000$ & $39{,}840$ & $0.39$ & $83.6$ & $3.10$ & $7.25$ & $94.3$ & $615$ \\
Erd\H{o}s-R\'{e}nyi & $4{,}000$ & $80{,}007$ & $4.57$ & $1{,}456$ & $27.78$ & $59.47$ & $1{,}548$ & $1{,}903$ \\
\midrule
Caltech36 & $762$ & $16{,}651$ & $0.03$ & $2.0$ & $0.18$ & $0.38$ & $2.6$ & $171$ \\
Reed98 & $962$ & $18{,}812$ & $0.05$ & $4.9$ & $0.37$ & $0.78$ & $6.2$ & $230$ \\
Haverford76 & $1{,}446$ & $59{,}589$ & $0.14$ & $25.6$ & $1.30$ & $2.72$ & $29.8$ & $358$ \\
Simmons81 & $1{,}510$ & $32{,}984$ & $0.16$ & $30.3$ & $1.49$ & $3.10$ & $35.1$ & $390$ \\
Swarthmore42 & $1{,}657$ & $61{,}049$ & $0.21$ & $43.6$ & $1.99$ & $4.12$ & $49.9$ & $444$ \\
Amherst41 & $2{,}235$ & $90{,}954$ & $0.59$ & $145.7$ & $4.91$ & $10.41$ & $161.6$ & $729$ \\
Bowdoin47 & $2{,}250$ & $84{,}386$ & $0.59$ & $147.0$ & $5.03$ & $10.62$ & $163.2$ & $729$ \\
Hamilton46 & $2{,}312$ & $96{,}393$ & $0.66$ & $163.4$ & $5.46$ & $11.52$ & $181.0$ & $765$ \\
\midrule
Email-Eu-core & $986$ & $16{,}064$ & $0.05$ & $5.6$ & $0.42$ & $0.84$ & $6.9$ & $231$ \\
\bottomrule
\end{tabular}
\end{table}

As the table shows, forming $M$ is the bottleneck, with time growing as $N^{4}$. The per-source stages (``solve'' and ``reconstruct'') are a small fraction of the setup. As anticipated, the running times for empirical networks match those of Erd\H{o}s-R\'{e}nyi graphs of the same size despite having very different densities. Memory grows as $N^{2}$ and is still only around $2$ GB when $N=4{,}000$. For comparison, a direct sparse solver (\texttt{scipy.sparse.linalg.spsolve}) took $27$ seconds and $1.2$ GB at $N=100$ and $366$ seconds and $4.5$ GB at $N=150$, whereas Algorithm~\ref{alg:meeting} takes well under a second at these sizes (and much less memory). Iterative solvers mitigate some of the time and space requirements, but their cost depends on the density of the graph, a tolerance, and the spectral gap.

\subsection{Vertex-transitive graphs}\label{sec:vertex_transitive}
On a vertex-transitive graph, the automorphism group of the graph acts transitively on the nodes, meaning a single walker cannot tell where they are on the graph at any given time. If $\varphi$ is an automorphism of the graph, then for meeting times, we have
\begin{align}
\tau_{\varphi\left(i\right)\varphi\left(j\right)} &= \begin{cases}
	0 & \varphi\left(i\right) =\varphi\left(j\right) , \\
	1 + \frac{1}{2}\sum_{k=1}^{N}P_{\varphi\left(i\right) k}\tau_{k\varphi\left(j\right)}+\frac{1}{2}\sum_{k=1}^{N}P_{\varphi\left(j\right) k}\tau_{\varphi\left(i\right) k} & \varphi\left(i\right)\neq\varphi\left(j\right) ,
\end{cases} \nonumber \\
&= \begin{cases}
	0 & i=j , \\
	1 + \frac{1}{2}\sum_{k=1}^{N}P_{ik}\tau_{\varphi\left(k\right)\varphi\left(j\right)}+\frac{1}{2}\sum_{k=1}^{N}P_{jk}\tau_{\varphi\left(i\right)\varphi\left(k\right)} & i\neq j ,
\end{cases}
\end{align}
from which we see that $\tau_{\varphi\left(i\right)\varphi\left(j\right)}=\tau_{ij}$ for all $i,j=1,\dots ,N$. Thus, the left-hand side of
\begin{align}
\tau - \frac{1}{2}P\tau - \frac{1}{2}\tau P^{\T} &= \mathbf{1}\mathbf{1}^{\T} + D
\end{align}
is invariant under $\varphi$, which means the right-hand side must be as well, giving a single unknown that determines $D$. This symmetry drops the cost of Algorithm~\ref{alg:meeting} to $O\left(N^{3}\right)$. This applies to the complete graph, cycle, torus, hypercube, and any graph whose automorphism group acts transitively on the vertices. However, regularity alone is not enough. For example, a simple calculation shows that $d_{i}$ is not constant on the Frucht graph \citep{frucht:CJM:1949}, which is a $3$-regular graph whose automorphism group is trivial.

Similar arguments apply to graphs with fewer symmetries. For example, if the automorphism group of the graph has $m$ orbits on the vertices, then $d$ is constant on each orbit and \eq{correction_constraint} has $m+1$ distinct unknowns. However, for this property to be useful, the orbits must be known in advance, which itself is nontrivial for a graph without evident global symmetry.

\section{Applications to evolutionary dynamics}\label{sec:evol}

\subsection{Fixation probabilities}
Following \citet{mcavoy:NHB:2020}, consider an additive evolutionary game on a weighted, connected graph in which a producer (type $A$) at vertex $i$ donates a benefit $B_{ij}$ to each neighbour $j$ at a cost $C_{ij}$ to itself, while a non-producer (type $B$) donates nothing and pays nothing. Writing $\bm{x}\in\left\{0,1\right\}^{N}$ for the population state, where $x_{i}=1$ means $i$ is a producer and $x_{i}=0$ means $i$ is a non-producer, we see that the payoff to individual $i$ is
\begin{align}
U_{i}\left(\bm{x}\right) &= \sum_{j=1}^{N}\left(-x_{i}C_{ij}+x_{j}B_{ji}\right) . \label{eq:additive_payoff}
\end{align}
Payoff gets translated into a positive reproduction propensity, $e^{\delta U_{i}\left(\bm{x}\right)}$. At each time step, an individual is chosen uniformly at random from the population for death, and the neighbors compete to reproduce and fill the vacancy with probability proportional to $e^{\delta U_{i}\left(\bm{x}\right)}$. Under death-Birth updating with uniform mutant appearance, producers are favored over non-producers under weak selection ($\delta \to 0$) whenever the inequality
\begin{align}
\sum_{i,\ell =1}^{N}\pi_{i}\left(-\left(1-\tau_{ii}\right) C_{i\ell} + \left(1-\tau_{i\ell}\right) B_{\ell i}\right) &> \sum_{i,j,\ell =1}^{N}\pi_{i}P_{ij}^{\left(2\right)}\left(-\left(1-\tau_{ij}\right) C_{j\ell}+\left(1-\tau_{i\ell}\right) B_{\ell j}\right) \label{eq:sel_cond}
\end{align}
holds, where $P_{ij}^{\left(n\right)}\coloneqq\left(P^{n}\right)_{ij}$ is the $n$-step transition probability and $\tau_{ij}$ is the expected meeting time of the lazy pair walk of \eq{meeting} \citep{allen:Nature:2017,mcavoy:NHB:2020}. By ``favored,'' we mean in the sense of \citet{mcavoy:NHB:2020}, meaning if $\rho_{A}$ is the fixation probability of a single producer in a population of non-producers and $\rho_{B}$ is the fixation probability of a single non-producer in a population of producers, then increasing $\delta$ from $0$ to a small, positive value increases the value of $\rho_{A}-\rho_{B}$. Algorithm~\ref{alg:meeting} computes all $\binom{N}{2}$ meeting times in $O\left(N^{4}\right)$, after which \eq{sel_cond} can be evaluated in $O\left(N^{3}\right)$ for any choice of $\left(B_{ij},C_{ij}\right)$. The expensive step depends only on the graph, not on the social good. Under weak selection, it therefore needs to be done only once per graph, however many social goods are compared.

\begin{figure}
\centering
\includegraphics[width=0.75\textwidth]{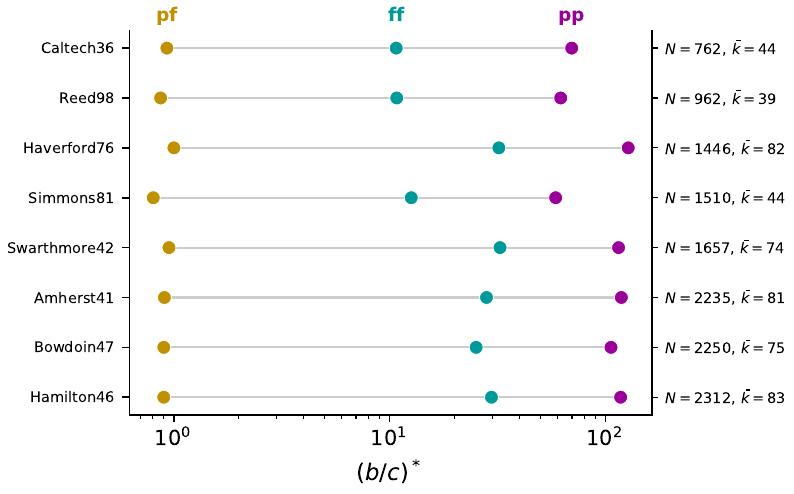}
\caption{Critical benefit-to-cost ratios under death-Birth updating on eight empirical social networks from Facebook \citep{rossi:AAAI:2015}. Three social-good schemes are depicted. Here, the sizes of the networks range from $N=762$ to $N=2{,}312$, and the average degree ranges from $39$ to $83$. Computing all meeting times took between $3$ seconds and around $3$ minutes per network on a laptop (see Table~\ref{tab:benchmarks}).}\label{fig:facebook_networks}
\end{figure}

\begin{figure}
\centering
\includegraphics[width=0.95\textwidth]{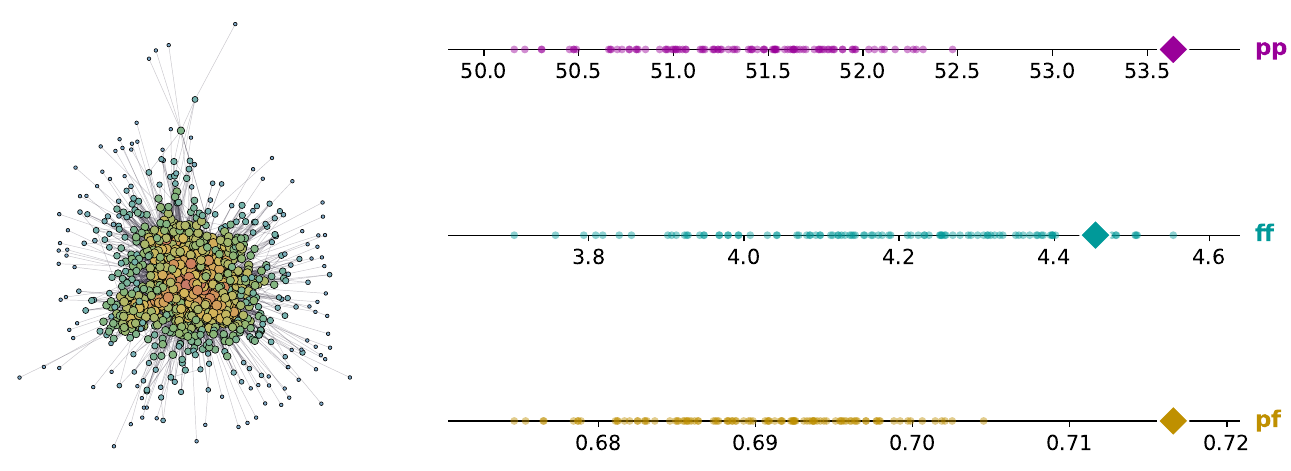}
\caption{Critical benefit-to-cost ratios under death-Birth updating on the Email-Eu-core social network of \citet{leskovec:ACM:2007}. Here, $N=986$, $\left| E\right| =16{,}064$, and the average degree is $\approx 33$. For the three social-good schemes, the empirical $\left(b/c\right)^{\ast}$ on the network (diamond) is plotted against one hundred independent degree-preserving rewirings of it (depicted as dots, with the rewirings obtained via $10\left| E\right|$ double-edge swaps \citep[see][]{maslov:Science:2002}). This result shows that the threshold for producers to thrive is generically \emph{higher} on this empirical network than on networks of comparable degree sequence. Computing all meeting times took about $7$ seconds for this network, for each rewiring (see Table~\ref{tab:benchmarks}).}\label{fig:email_network}
\end{figure}

As illustrations, Figures~\ref{fig:facebook_networks}~and~\ref{fig:email_network} compute the critical benefit-to-cost ratios on several empirical networks. Three kinds of social goods of \citet{mcavoy:NHB:2020} are considered, namely \emph{pp} (proportional benefits, proportional costs), \emph{ff} (fixed benefits, fixed costs), and \emph{pf} (proportional benefits, fixed costs). In the three cases, we have $\left(B_{ij},C_{ij}\right) =\left(bW_{ij},cW_{ij}\right)$ (pp goods), $\left(B_{ij},C_{ij}\right) =\left(bW_{ij}/w_{i},cW_{ij}/w_{i}\right)$ (ff goods), and $\left(B_{ij},C_{ij}\right) =\left(bW_{ij},cW_{ij}/w_{i}\right)$ (pf goods). Figure~\ref{fig:facebook_networks} shows the critical ratio for $b/c$, which is the value beyond which producers of social goods are favored by selection, on eight empirical Facebook networks. Figure~\ref{fig:email_network} shows the Email-Eu-core social network of \citet{leskovec:ACM:2007}. For each scheme in the email network, the empirical $\left(b/c\right)^{\ast}$ sits about $5\%$ above all $100$ degree-preserving rewirings of the graph. We show these examples only to illustrate the kind of question Algorithm~\ref{alg:meeting} makes accessible at $N\sim 10^{3}$, regardless of network density.

\subsection{Identity by state and mutation}
The structure-coefficient theorem of \citet{tarnita:JTB:2009} for weak selection with recurrent mutation (probability $u>0$) is built on the pairwise probability of identity by state $\phi_{ij}$, which under death-Birth updating satisfies $\phi_{ii}=1$ for all $i=1,\dots ,N$ and
\begin{align}
\phi_{ij} = u\frac{1}{2} + \left(1-u\right)\frac{1}{2}\sum_{k=1}^{N}P_{ik}\phi_{kj}+\left(1-u\right)\frac{1}{2}\sum_{k=1}^{N}P_{jk}\phi_{ik} \label{eq:phi}
\end{align}
for $i\neq j$ \citep{mcavoy:PNAS:2022}. This is the prescribed-diagonal Poisson equation (\eq{poisson}) with damping factor $\left(1-u\right) /2$ in place of $1/2$, constant off-diagonal source ($u/2$), and $h_{i}=1$ for $i=1,\dots ,N$. Algorithm~\ref{alg:meeting} applies with minor modifications, computing all $\binom{N}{2}$ identity-by-state probabilities in $O\left(N^{4}\right)$. Since pairwise identity-by-state probabilities determine the structure coefficients under weak-selection (see Box $1$ of \citet{mcavoy:PNAS:2022}), this also gives an $O\left(N^{4}\right)$ algorithm for computing mean trait frequencies under recurrent mutation on arbitrary weighted graphs.

\section{Discussion}\label{sec:discussion}
The core of this argument is the diagonal correction, since the rest follows from structure exploited by the Bartels-Stewart algorithm. Here, the diagonal correction is unknown, and determining the value of this unknown determines the time-complexity of solving the original system. In this sense, the argument follows a technique used in \citep{mcavoy:JMB:2021}, where a large linear system could be replaced by a smaller one, based on expressing unknowns in terms of recurrences. Here, too, we reduce an unknown (the diagonal correction) to a linear system (\eq{correction_constraint}). However, the bottleneck of the complexity reported here is not in solving this system but in \emph{forming} it in the first place.

The numerical stability of Algorithm~\ref{alg:meeting} is essentially the same as that of the standard Bartels-Stewart algorithm. The divisors $1-\frac{1}{2}\lambda_{i}-\frac{1}{2}\lambda_{j}$ in \eq{HV_entry} are the eigenvalues of the Sylvester operator and govern the conditioning of both methods. The diagonal correction involves an additional linear system without any new source of ill-conditioning. Since $\lambda_{1}=1$, the smallest divisor among $\left(i,j\right)\neq\left(1,1\right)$ is at $\left(i,j\right) =\left(1,2\right)$, which is $\frac{1}{2}\left(1-\lambda_{2}\right)$. Therefore, conditioning is poor precisely for slowly mixing graphs, such as cycles (where $1-\lambda_{2}\sim N^{-2}$) or dense clusters joined by only a small number of edges.

We note that laziness of the pair walk, where exactly one walker moves at a time, is crucial. When both walkers move simultaneously, the pair-walk transition matrix is the Kronecker \emph{product} $P\otimes P$ \citep{george:arXiv:2018}, for which the Sylvester decomposition does not apply. While the simultaneous-move version may be of independent interest, our motivation here comes not from random walks themselves but rather from how they arise in the study of evolutionary dynamics. For the latter, the lazy pair walk is what emerges naturally \citep{allen:Nature:2017,mcavoy:JMB:2021}, and we hope that the algorithm presented here will be practically useful for studying questions related to the evolutionary dynamics of social behaviors in larger, heterogeneous populations.

\section*{Acknowledgment}
Thanks to Yoichiro Mori for an insightful conversation about the Sylvester equation several years ago.

\section*{Code availability}
Supplementary code is available at \texttt{https://github.com/alexmcavoy/meeting-times}, which contains an implementation of the algorithm for calculating critical ratios, code to generate the figures, and benchmarks. Claude Code (Opus 4.6 and Fable 5.1) was used to check and streamline the implementation, as well as to develop the benchmarks. All responsibility remains with the author.

\end{document}